\documentclass[a4paper, amsfonts, amssymb, amsmath, reprint, showkeys, nofootinbib, twoside]{revtex4-1}

\usepackage{amsmath,amssymb,bm}
\usepackage{graphicx}
\usepackage{subcaption}
\usepackage{physics}
\usepackage{hyperref}
\usepackage{xcolor}
\usepackage{microtype}

\begin{document}

\title{Stochastic systems with Bose--Hubbard interactions:\\
Effects of bias on particles {on a} random comb}

\author{Swastik Majumder${}^{1,2}$}
\email{majumderswastik09@gmail.com}

\author{Mustansir Barma${}^{1}$}
\email{barma@tifrh.res.in}

\affiliation{${}^{1}$ Tata Institute of Fundamental Research, Hyderabad,\\
36/P, Gopanpally Village, Serilingampally Mandal, Hyderabad, Telangana 500046, India}
\affiliation{${}^{2}$ Indian Institute of Science Education and Research Kolkata,\\
Campus Rd, Mohanpur, Haringhata Farm, West Bengal 741246, India}

\date{\today}

\begin{abstract}
We study stochastic transport of interacting particles on a disordered network described by the
random comb geometry. The model is defined on a one-dimensional backbone from which branches of
random lengths emanate, providing a minimal model of percolation networks beyond the critical
percolation probability. The dynamics obeys local detailed balance with respect to a Bose-Hubbard
Hamiltonian containing both an external bias and on-site repulsion. This choice yields an analytically
tractable steady state through a mapping to the zero-range-process. We compute the backbone current,
branch density profiles, and macroscopic drift velocity, and analyze how bias and interactions compete
to shape transport. The backbone current increases monotonically with density, while the drift
velocity displays a non-monotonic dependence on the external field, remaining finite for any nonzero
bias, in contrast to the vanishing drift velocity of noninteracting particles beyond a threshold bias. Density profiles along branches
exhibit stepwise plateaus governed by the ratio of interaction to bias energy. These results highlight
how repulsive interactions suppress trapping and restore transport in disordered geometries, bridging
earlier studies of field induced drift in random networks with the physics of disordered Bose-Hubbard
systems.

\end{abstract}

\maketitle

\section{Introduction}

Transport in disordered or complex networks is a central topic in nonequilibrium statistical physics~\cite{Havlin1987,Bouchaud1990}. In particular,
quenched disorder leading to reduced connectivity can combine with external fields to produce nontrivial steady 
states characterized by anomalous drift and dispersion, long-lived trapping, and slow relaxation. 

Simplified lattice models have proved to be very valuable in understanding such effects. Studies of biased diffusion of
non-interacting particles on percolation clusters reveal that as the bias is increased, particles experience enhanced 
trapping in branches, causing the backbone drift velocity to vanish beyond a threshold field~\cite{Barma1983, Dhar1998}. A particularly transparent geometry which captures the essential interplay between directed motion and trapping  is provided 
by the random comb (Fig.~\ref{fig:random_comb})\cite{White1984}, which consists of a one-dimensional backbone from which branches of random 
lengths emanate, mimicking and simplifying the geometry of the infinite cluster in percolation.  On the random comb,
 the drift velocity and dispersion can be found exactly, allowing the determination of the thresholds of bias beyond which
drift and dispersion become anomalous\cite{White1984,Bunde1986,Pottier1995,Kotak2022,Balakrishnan1995}.

When inter-particle interactions are introduced, the qualitative nature of transport changes significantly. Exclusion effects
which account for hard-core interactions between particles restore a finite current at all values of the bias 
by preventing overcrowding in traps, as shown for lattice gases on disordered networks~\cite{Iyer2025, Ramaswamy1987, SG}. At the same time, a tagged particle  in a branch spends
an extremely  long time there, resulting in an escape probability that decays slower than any power law~\cite{Iyer2025}.

In this work, we study a model of biased diffusion in a model which allows for multiple occupancy, but with repulsive Bose–Hubbard-type on-site interactions on a random comb. An earlier 
paper~\cite{Majumder2025} analyzed the system in the absence of disorder, using a stochastic dynamical rule which obeys local detailed 
balance with respect to the interaction Hamiltonian, but for which the steady state could not be found exactly. In this work, 
the hopping rates of particles are chosen differently: they continue to obey local detailed balance, but reduce to the zero range process 
(ZRP)~\cite{Evans2005}, enabling the exact steady state to be found. This enables the derivation of explicit forms for the backbone current, 
branch density profiles, and overall drift velocity, showing that even though non-interacting systems can show complete trapping, interactions 
preserve a finite current for all values of the bias. While the steady states on the periodic backbone are quite different in~\cite{Majumder2025} and the present 
work, free boundary conditions lead to an identical equilibrium state for both cases.

In Section II, we define the random comb and the stochastic moves of particles on it. Since branches of the random comb carry no current, their 
presence does not affect the steady state on the periodic backbone ring. Thus in Section III, we consider particles on the backbone alone, then
particles in a branch in Section IV and finally on the full random comb in Section V. The conclusions are presented in Section VI.

\section{Model} 
\label{sec:random}
\begin{figure}[htbp]
    \centering
    \includegraphics[width=0.4\textwidth]{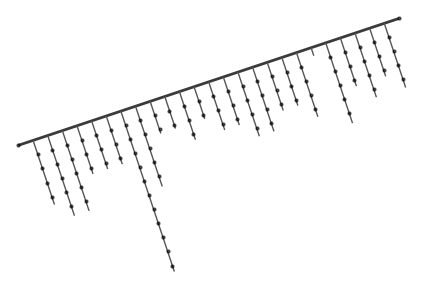}
    \caption{Random comb structure with a tilted backbone and branches.}
    \label{fig:random_comb}
\end{figure}

We consider the problem on a random comb, which consists of a backbone, with branches of random lengths~\cite{Havlin1987} attached to each site in the backbone. We assign a distribution function for branches of length $l$~\cite{SG}.\begin{align}
    P(l) = \left(1 - e^{-a/\zeta}\right) e^{-a l / \zeta}.
\end{align}
Here $a$ is the lattice spacing and, $\zeta$ is the correlation length. The exponential decay of branch lengths mimics that in the percolation problem. A typical random comb is depicted in Fig.~\ref{fig:random_comb}.

Since the random comb is a relatively simple structure which embodies some key features of random media, a number of studies involving transport have been performed  on it earlier. Non-interacting particles performing unbiased random walks exhibit a finite, though reduced, diffusion constant on large scales~\cite{Aslangul1994}. The introduction of bias leads to new effects coming from a competition between drift along the backbone and strong trapping in branches~\cite{White1984, Bunde1986, Pottier1995, Kotak2022, Balakrishnan1995}. This leads to a threshold value of the bias beyond which the drift velocity vanishes, and a lower value at which the dispersion becomes anomalous~\cite{Kotak2022}. 

The introduction of hard core interactions changes the nature of transport significantly, as a particle on the backbone usually sees other particles in the branch, which limit its ingress~\cite{Ramaswamy1987, Iyer2025, SG}. At the same time, a tagged particle  in a branch spends an extremely  long time localized there, resulting in an escape probability that decays slower than any power law~\cite{Iyer2025}.

Bose-Hubbard interactions of the type under study here embody interparticle repulsion, but not as strongly as hard core exclusion. Such interactions have been  used to describe bosons at low temperatures in optical trap~\cite{Duchon2013}. Further, the effects of tilt (i.e. bias)~\cite{Sachdev2002, Lake2022, Lake2023} and quenched disorder~\cite{Choi2016, DeMarco2017, Scarola2017, DeMarco2025} have also been explored in such a setting. At higher temperatures where quantum effects are negligible, particles are expected to move stochastically, as in this work. Our results, on the combined effect of bias and disorder in a classical stochastic setting (keeping the Bose-Hubbard interaction intact) may provide interesting pointers for the dynamics of bosons in the non-degenerate regime at higher temperatures.

Now let us turn to the rules which govern the stochastic dynamics of particles on the random comb. The rates satisfy a local detailed balance
condition, but differ from those used in~\cite{Majumder2025}. 

At site \(i\), there is an on-site energy \(\mathcal{E}(i)\) with contributions from the external field \(E\), which leads
to biased hopping, and from the repulsive Bose--Hubbard interaction \(U\) amongst particles on the site.  

\begin{align}
\mathcal{E}(i)
&= - E a\, i\, n_i + \frac{U}{2} n_i (n_i - 1),
\label{ener}
\end{align}

A single hop of a particle from site \(i+1\) to site \(i\) leads to a transition from configuration
\[
\mathcal{C} = (n_1, \ldots, n_i, n_{i+1}, \ldots, n_L),
\]
to a new configuration
\[
\mathcal{C}' = (n_1, \ldots, n_i - 1, n_{i+1} + 1, \ldots, n_L).
\]

The hopping rates are chosen so as to ensure that local detailed balance is maintained with respect to
$
E(\mathcal{C}) = \sum_k \mathcal{E}(k),
$
so that
\begin{align}
\frac{
W_1\bigl((n_i, n_{i+1}) \to (n_i - 1, n_{i+1} + 1)\bigr)
}{
W_2\bigl((n_i - 1, n_{i+1} + 1) \to (n_i, n_{i+1})\bigr)
}
&= e^{-\beta \left[ E(C') - E(C) \right]}.
\label{hopp}\end{align}

A choice of hopping rates (using a notation similar to that in~\cite{Majumder2025}) that satisfies the above local detailed balance condition in Eq.~\eqref{hopp} is: 
\begin{align}
W_1(n_i, n_{i+1})
&= \exp\!\left[ 0.5\beta E a + \beta U (n_i - 1)\,\right]\theta(n_i) , \\
W_2(n_i, n_{i+1})
&= \exp\!\left[ -0.5\beta E a + \beta U (n_{i+1} - 1)\, \right]\theta(n_{i+1}).
\label{rates}\end{align}

These are the rates we use in this study which are different from those used in~\cite{Majumder2025}. Note that the rates fall into the class of rates covered by the zero-range process (ZRP),
which facilitates the subsequent analysis.

\section{The periodic backbone}
Since the stochastic moves of particles are biased, the backbone with periodic boundary conditions carries a current. Its value is unaffected by the branches, since the boundary condition at the terminus of every branch ensures that there is no current in branches.  Thus we begin by considering particles on the backbone alone.

The dynamics on the backbone is described by the ZRP~\cite{Evans2005} with the rates specified in Eq.~\eqref{rates}. In the grand canonical ensemble, the probability of a given configuration follows a {product measure form}, given by:
\begin{equation}
    P(\mathcal{C}) = \prod_{i=1}^{L} z^{n_i} f(n_i),
\end{equation}
where \( f(m) \) is defined as
\begin{equation}
    f(m) = \prod_{n=1}^{m} \frac{1}{e^{\beta U (n-1)}} = \exp\left(-\frac{\beta U}{2} m (m-1)\right), f(0) = 1.
\end{equation}
The steady-state current \( j_B \) is then given by
\begin{equation}
    j_B = \left( e^{\frac{\beta E a}{2}} - e^{-\frac{\beta E a}{2}} \right) z,
    \label{eq:jb}
\end{equation}
where \( z \) is the fugacity. The fugacity is related to the backbone density \( \rho_0 \) through the following expression:
\begin{equation}
    \rho_0 = z \frac{F'(z)}{F(z)},
    \label{eq:rho0}
\end{equation}
where \( F(z) \) is defined as
\begin{equation}
    F(z) = \sum_{m=0}^{\infty} z^m f(m).
    \label{ref:eqF(m)}
\end{equation}
\renewcommand{\figurename}{FIG.}

The density \( \rho_0 \) in the grand canonical ensemble can then be written as
\begin{equation}
    \rho_0 = \frac{1}{\mathcal{Z}_{0}(z,\beta U)}\sum_{j=0}^{\infty} j z^j \exp\left( -\beta U \frac{j(j-1)}{2} \right).
\end{equation}
Using the relations in Eqs.~\eqref{eq:jb}, \eqref{eq:rho0}, and \eqref{ref:eqF(m)}, we can now obtain \( j_B \) as a function of \( \rho_0 \), which is a monotonically increasing function of the backbone density \( \rho_0 \). However, there are no signatures of quasi-ASEP-like behavior or non-monotonic correlations with respect to interaction strength like in~\cite{Majumder2025}. This absence can be attributed to the fact that such effects are most pronounced in the totally asymmetric limit, while in the present framework, a totally asymmetric scenario corresponds to an infinite biasing strength, effectively rendering the on-site repulsion irrelevant.

\section{The branches}
{Next, let us consider the steady state profile profile with a single branch, which consists of{ a linear chain with {$l+1$ sites labeled by $k=0,1,2,\dots,l$. The density at site $0$ (which lies on the backbone) is specified to be $\rho_0$ on the first site}, and open boundary conditions are applied at site $l$.  Below, we describe the density profiles in a branch, and the current carried in the backbone when the density is $\rho_0$.}}

In the steady state, there is no current in the branch, and the particles within this branch are described by the Hamiltonian in Eq.~\eqref{ener}. Introducing a fugacity $z$ in the branch, we obtain the density at each site in the branch 
  \begin{equation}
      \rho_k=\frac{1}{\mathcal{Z}_{k}(z,\beta U)}\sum_{n=0}^{\infty} nz^nexp\left(\beta Eakn-\beta U \frac{n(n-1)}{2}\right)\label{rhok}.
  \end{equation}
 {Due to the bias, the particle density increases toward higher-index sites.} Eq.~\eqref{rhok} can be well approximated by the{  Euler-Maclaurin} formula for numerical calculations.
\begin{figure}[h!]
    \centering
    \begin{subfigure}[b]{0.48\textwidth}
        \includegraphics[width=\textwidth]{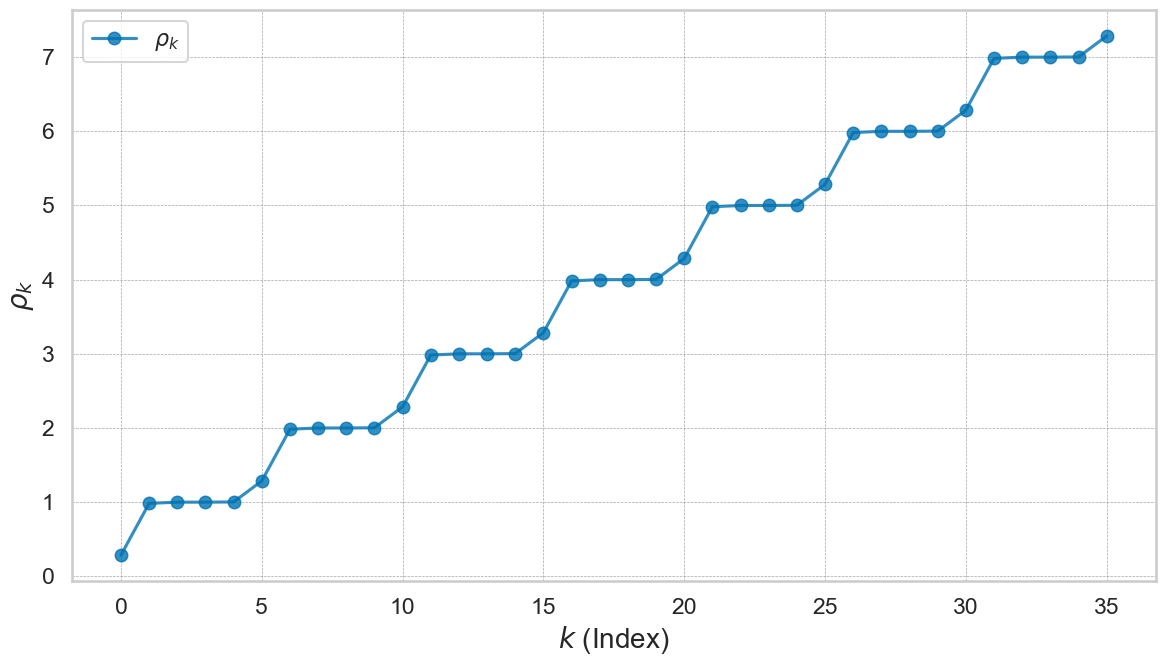}
       \caption{$\rho_k$ vs $k$, where $k$ is the lattice index of the branch for $\beta=5$ with the ratio of repulsion versus biasing term being $\frac{ U}{Ea}=5$, and fugacity $z=0.4$.}
     \label{fig:rhok1}
    \end{subfigure}
    \hfill
    \begin{subfigure}[b]{0.48\textwidth}
        \includegraphics[width=\textwidth]{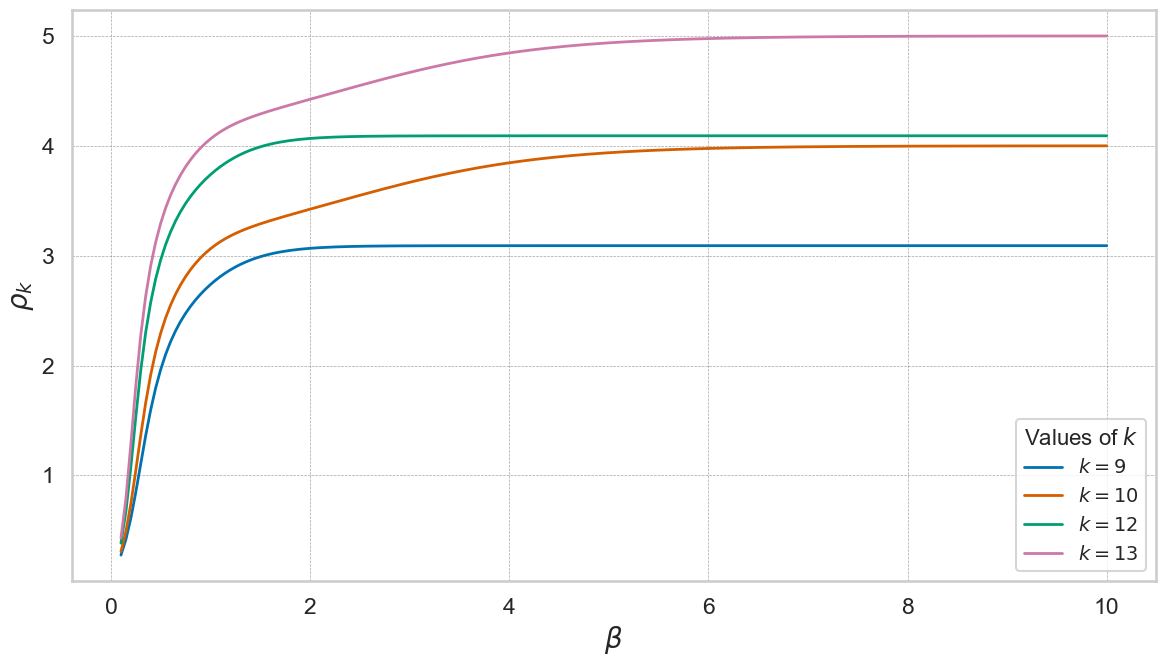}
         \caption{$\rho_k$ vs $\beta$, with the ratio of repulsion versus biasing term as, $\frac{U}{Ea}=3$, and fugacity $z=0.1$. Different curves indicate the lattice indices of the branch.}
        \label{fig:rhok2}
    \end{subfigure}
    \caption{Variation of average density $\rho_k$ as a function of branch depth $k$ and $\beta U$. }
    \label{figa}
\end{figure}

  The plot of \(\rho_k\) versus \(\beta\) in Fig.~\ref{figa} exhibits a periodic modulation governed by the ratio of repulsion to bias. This modulation determines the saturation values of \(\rho_k\), with \(k=9\) saturating at 3, \(k=10\) at 4, and so forth. The periodicity reflects the interplay between the energy contribution associated with \(k\) and the imposed bias, resulting in distinct plateaus aligned with the modulation period.

For \(U < Ea\), the average density increases monotonically and approximately linearly as one moves deeper into the branch. In contrast, for \(U > Ea\) (as shown), the density exhibits a stepwise increase, with each step characterized by unique features discussed below.

At very low temperatures, the average density at each site takes integer values. When the ratio \(\frac{U}{Ea}\) is an integer greater than one, the length of each step corresponds exactly to this ratio, as confirmed by energy minimization. If \(\frac{U}{Ea}\) is greater than one but not an integer, the step length can be approximated by rounding the ratio, consistent with energy minimization results.

\section{The random comb}
We now consider the problem on a random comb, which consists of a backbone, with branches of arbitrary lengths~\cite{White1984} attached to each site in the backbone. We assign a distribution function for having branches of length $l$
\begin{equation}
      P(l)=(1 - e^{-a / \zeta}) e^{-a l / \zeta}.\label{poss}
\end{equation}
Here $a$ is the lattice spacing and, $\zeta$ is the correlation length. The exponential decay of branch lengths mimics that in the percolation problem. 
A typical random comb is depicted in Fig.~\ref{fig:random_comb}.

\begin{figure}[h!]
    \centering
    \begin{subfigure}[b]{0.48\textwidth}
        \includegraphics[width=\textwidth]{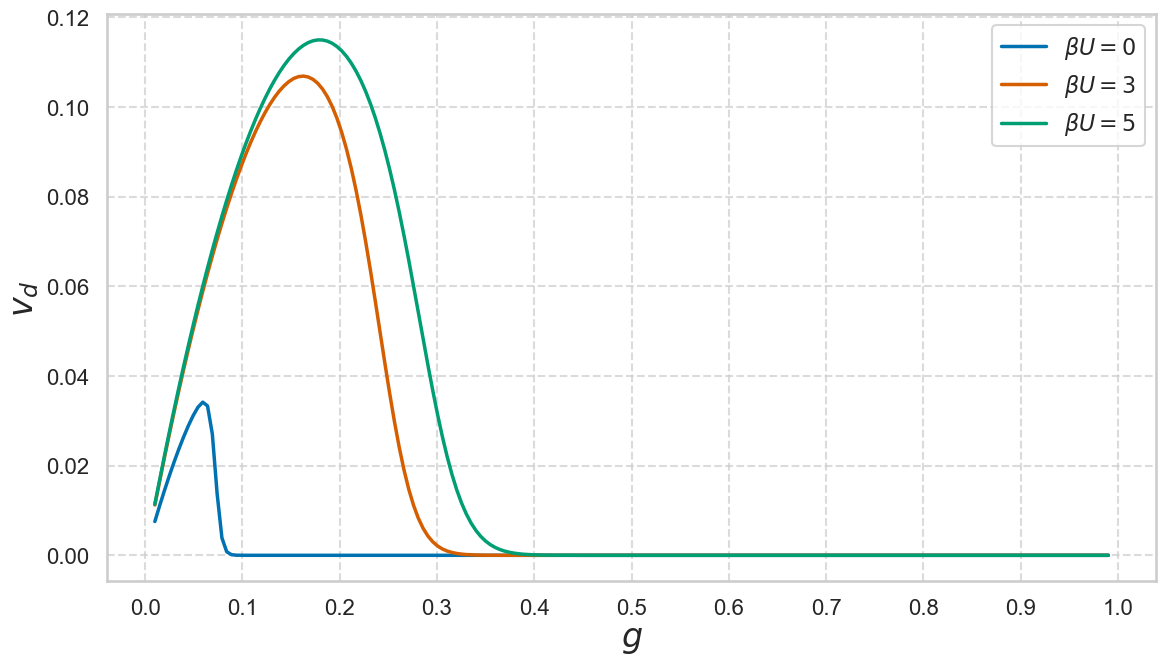}
       \caption{Drift velocity $v_d$ versus asymmetry parameter \(g\) at different values of $\beta U$. Other parameters: $a=1,\zeta=1.5,z=0.2$.}
        \label{vdg}
    \end{subfigure}
    \hfill
    \begin{subfigure}[b]{0.48\textwidth}
        \includegraphics[width=\textwidth]{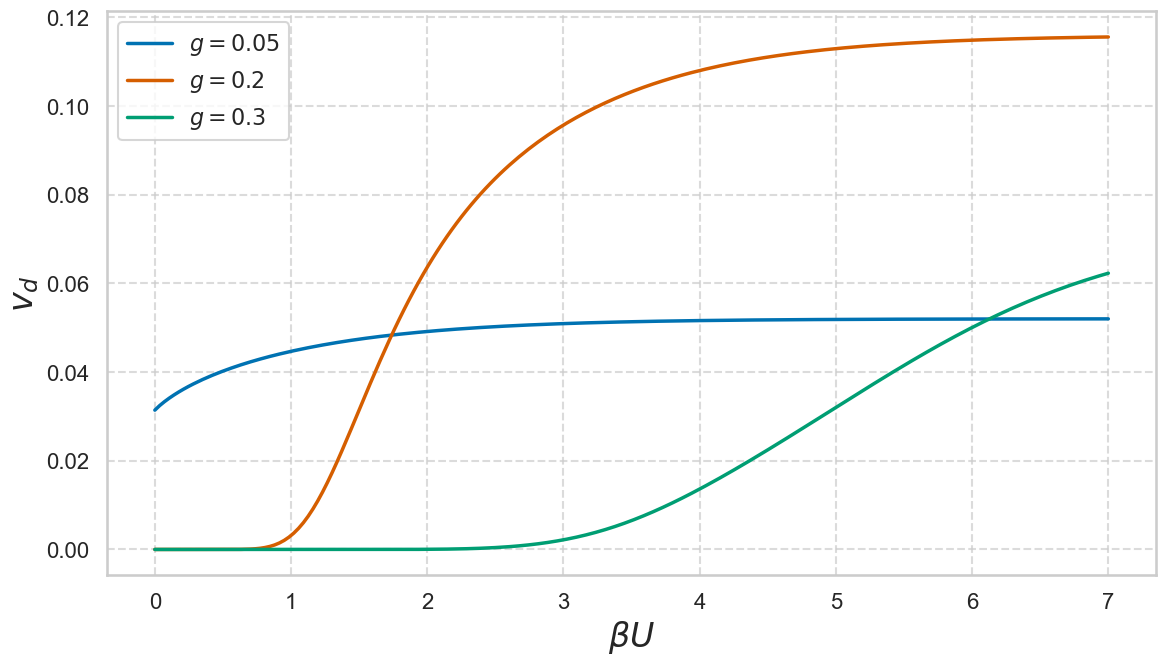}
         \caption{Drift velocity $v_d$ versus interaction strength \(\beta U\) at different values of $g$. Other parameters: $a=1,\zeta=1.5,z=0.2$.}
        \label{vdu}
    \end{subfigure}
    \caption{Dependence on the drift velocity $v_d$ as a function of  bias $g$ and interaction strength $\beta U$. }
    \label{rbcx}
\end{figure}

In the limit of a large backbone, that is the number of sites in the backbone going to $\infty$, it has been shown in~\cite{White1984, Barma1983,Ramaswamy1987} that the {macroscopic} drift velocity $v_d$ in the overall comb is given by, 
\begin{equation}
    v_d=\frac{j_B}{\bar{n}},
\end{equation}
where \(j_B\) is the backbone current and \(\bar{n}\) is the average number of particles in a branch. For a given value of \(z\), \(j_B\) and \(\bar{n}\) are computed using Eq.~\eqref{eq:jb} and Eq.~\eqref{poss}, together with Eq.~\eqref{rhok}, which gives the occupancy of each site.

The drift velocity \(v_d\) exhibits a non-monotonic dependence on the asymmetry parameter \(g\), which is defined through,
\begin{align}
    \frac{1+g}{1-g}=e^{\beta Ea}
\end{align}This behavior arises because the underlying {density profiles} \( \bar{n} \) are not simple functions of \( g \) alone but also depend crucially on the product \( \beta U \).
The overall behavior of \( v_d \) as a function of \( g \) and \( U \) results from a {competition between asymmetry-driven transport} and {potential-dominated localization} effects as shown in Fig.~\ref{rbcx}. The interplay of two factors underlies the non-monotonic structure observed in the drift velocity. However, the growth of $\rho_k$ with the depth of the branch is slower than linear as opposed to an exponential growth. As a result, the trapping behavior of particles as seen in \cite{White1984,Barma1983} is mitigated and the drift velocity does not vanish for any finite strength of the bias.

We can also consider the random system under free boundary conditions, which can be experimentally realized in disordered platforms~\cite{DeMarco2017,Scarola2017}. Under such conditions with suitable values of $U$ and $Ea$, the particle density along the backbone also exhibits a stepwise growth, while the branch densities display corresponding step-like growth patterns which is governed by the local backbone density at the attachment site.
\section{Conclusion}
We have investigated stochastic transport on a random
comb in the presence of on-site Bose–Hubbard
interactions and an external bias. By imposing local
detailed balance with respect to the interaction Hamiltonian,
the dynamics admitted an exact zero-range-process
representation, allowing analytical determination of the
steady-state properties, both on the backbone and in the 
branches. Our study reveals that repulsive
interactions play a crucial role in mitigating field-induced
trapping and restoring transport in the disordered network.
While non-interacting systems on the random comb exhibit
a vanishing drift velocity beyond a threshold bias, the
inclusion of on-site repulsion maintains a finite backbone
current and a non-vanishing macroscopic drift velocity
for any finite bias. The density profiles along branches
develop stepwise plateaus, indicating discrete trapping 
thresholds determined by the ratio of interaction to bias energy.

An additional dynamical observable of interest is the average
time taken by a particle to escape from the bottom of a branch and
return to the backbone. In the non-interacting case, this escape
time grows exponentially with the length of the branch $l$~\cite{Barma1983, Dhar1998, White1984}. By contrast, the
escape time in the case of hard core particles grows as 
$exp( \beta E l^2/2)$ as $O(l)$~\cite{ Ramaswamy1987,Iyer2025, BarmaRamaswamy1986}.
In our study, the repulsive Bose-Hubbard interactions involved 
are softer, and the escaping particle can find its way to
the backbone by a sequence of steps, each of which requires
stepping on intermediate  occupied sites. A detailed study of these
processes and  their implications for the escape time
are left for future study.

We have seen that the competition between disorder, interactions and bias 
on the random comb gives rise to interesting effects such 
as a nonmonotonic current and density profiles with
pronounced multiple  steps. These results for a classical stochastic
model of biased diffusion with Bose-Hubbard interactions 
in a random network may help in developing the phenomenology
for the dynamics of interacting bosons in 
disordered optical lattices, in the nondegenerate regime at high temperature. 
Of course, it would be even more interesting to study interacting bosons on a 
random comb at low temperature, when quantum effects dominate.  Results 
for this  interesting open problem would help to shed light on the larger 
problem of quantum transport in disordered media.

\section{Acknowledgements} \label{sec:acknowledgements}
We thank Prof. Arti Garg for drawing our attention to the similarity between our model and Bose–Hubbard type interactions, and for insightful discussions on this topic. S.M. thanks TIFR Hyderabad for supporting this work during the Visiting Students Research Programme (VSRP) and subsequent visits as a visiting student. M.B. acknowledges the support of the Indian National Science Academy (INSA). We also acknowledge the support of the Department of Atomic Energy, Government of India, under Project Identification No. RTI4007. 
\bibliographystyle{unsrt}

\end{document}